\documentclass[journal, twocolumn]{IEEEtran}
\IEEEoverridecommandlockouts
\usepackage{amsmath,amssymb,amsfonts}
\usepackage{algorithmic}
\usepackage{bbm}
\usepackage{bm}
\usepackage{cite}
\usepackage{graphicx}
\usepackage{textcomp}
\usepackage{xcolor}

\usepackage{hyperref}
\hypersetup{hidelinks, 
colorlinks=true,
allcolors=black,
pdfstartview=Fit,
breaklinks=true}

\def\BibTeX{{\rm B\kern-.05em{\sc i\kern-.025em b}\kern-.08em
    T\kern-.1667em\lower.7ex\hbox{E}\kern-.125emX}}
\begin{document}

\bibliographystyle{IEEEtran}

\newcommand{\myincludegraphics}[2][width=8.5cm]{\includegraphics[#1]{#2}}

\title{Finite-Time Capacity: Making \\Exceed-Shannon Possible?}

\author{\IEEEauthorblockN{Jieao~Zhu, Zijian~Zhang, Zhongzhichao~Wan, and~Linglong~Dai}
	\thanks{
		All authors are with the Beijing National Research Center for Information Science and Technology (BNRist) as well as the Department of Electronic Engineering, Tsinghua University, Beijing 100084, China (E-mails: zja21, zhangzj20, wzzc20@mails.tsinghua.edu.cn; daill@tsinghua.edu.cn).
	
	This work was supported in part by the National Key Research and Development Program of China (Grant No.2020YFB1807201), in part by the National Natural Science Foundation of China (Grant No. 62031019), and in part by the European Commission through the H2020-MSCA-ITN META WIRELESS Research Project under Grant 956256.
	}
}

\maketitle

\begin{abstract}
Shannon-Hartley theorem can accurately calculate the channel capacity when the signal observation time is infinite. However, the calculation of finite-time capacity, which remains unknown, is essential for guiding the design of practical communication systems. In this paper, we investigate the capacity between two correlated Gaussian processes within a finite-time observation window. We first derive the finite-time capacity by providing a limit expression. Then we numerically compute the maximum transmission rate within a single finite-time window. We reveal that the number of bits transmitted per second within the finite-time window can exceed the classical Shannon capacity, which is called as the Exceed-Shannon phenomenon. Furthermore, we derive a finite-time capacity formula under a typical signal autocorrelation case by utilizing the Mercer expansion of trace class operators, and reveal the connection between the finite-time capacity problem and the operator theory. Finally, we analytically prove the existence of the Exceed-Shannon phenomenon in this typical case, and demonstrate the achievability of the finite-time capacity and its compatibility with the classical Shannon capacity.
\end{abstract}

\begin{IEEEkeywords}
Finite-time capacity, Exceed-Shannon, Mercer expansion, signal autocorrelation, operator theory.
\end{IEEEkeywords}

\newtheorem{theorem}{\bf Theorem}
\newtheorem{lemma}{\bf Lemma}
\newtheorem{remark}{Remark}

\section{Introduction}

The Shannon-Hartley theorem \cite{Shannon} has accurately revealed the fundamental theoretical limit of information transmission rate $C$, which is also called as the Shannon capacity, over a Gaussian waveform channel of a limited bandwidth $W$. The expression for Shannon capacity is $C=W\log\left(1+S/N\right)$, where $S$ and $N$ denote the signal power and the noise power, respectively. The derivation of Shannon-Hartley Theorem heavily depends on the Nyquist sampling principle \cite{Sampling1967Landau}. The Nyquist sampling principle, which is also named as the $2WT$ theorem \cite{Nyquist1928CertainTopics}, claims that one can only obtain $2WT + o(2WT)$ independent samples within an observation time window $T$ in a channel band-limited to $W$ \cite{BW1976Slepian}, where $o(\cdot)$ means higher-order infinitesimal, i.e., $o(2WT)/T\to 0, T\to \infty$.

Based on the Nyquist sampling principle, the Shannon capacity is derived by multiplying the capacity $1/2 \log (1+P / N)$ of a Gaussian symbol channel \cite[p.249]{Cover1999ElementsInfTheory} with $2WT+o(2WT)$ at first, and then dividing the result by $T$, finally letting $T\rightarrow \infty$. In the above derivation, $(2WT+o(2WT))/T$ is approximated by $2W$ in the final step to obtain the Shannon capacity. Note that this approximation only holds when $T \to \infty$. Therefore, the Shannon capacity only asymptotically holds as $T$ becomes sufficiently large. When $T$ is of finite value, the approximation fails to work. Thus, when the observation time $T$ is finite, i.e., the received signal can only be observed within a finite-time window $[0,T]$, Shannon-Hartley Theorem cannot be directly applied to calculate the capacity in a finite-time window. To the best of our knowledge, the evaluation of the finite-time capacity has not yet been investigated in the literature. One possible reason is that, most of the researchers mainly focused on how to approximate the Shannon capacity with advanced coding and modulation schemes. It is worth noting that any real-world communication systems transmit signals in a finite-time window, thus evaluating the finite-time capacity is of practical significance. \par 

In this paper, to fill in this gap, the finite-time capacity instead of the traditional infinite-time counterpart is analyzed, where we reveal and prove the existence of ``Exceed-Shannon'' phenomenon within a finite-time observation window\footnote{Simulation codes will be provided to reproduce the results presented in this paper: \href{http://oa.ee.tsinghua.edu.cn/dailinglong/publications/publications.html}{http://oa.ee.tsinghua.edu.cn/dailinglong/publications/publications.html}.}. Specifically, our contributions are summarized as follows:

\begin{itemize}
	\item We derive the capacity expressions within a finite-time observation window by using dense sampling and limiting methods. In this way, we can overcome the continuous difficulties that appear when analyzing the information contained in a continuous time interval. These finite-time capacity expressions make the analysis of finite-time capacity problems possible.
    \item We approximate the original continuous finite-time capacity expressions by discrete matrices, and conduct numerical experiments based on the discretized formulas. In the numerical results under a special setting, we reveal the ``Exceed-Shannon'' phenomenon\footnote{In fact, the finite-time ``Exceed-Shannon'' phenomenon revealed in this paper does not contradict the classical infinite-time Shannon-Hartley theorem, since new assumptions are considered. Specifically, in the Shannon-Hartley theorem, the sampling time is assumed to be infinitely long, while in this paper, the sampling takes place in a finite-time observation window. Similarly, although compressed sensing\cite{donoho2006compressedsensing} can achieve much lower sampling rate than the Nyquist sampling rate to perform accurate sparse signal reconstruction, it does not contradict the Nyquist sampling principle due to the new assumption of signal sparsity.}, i.e., the mutual information within a finite-time observation window exceeds the Shannon capacity.
    \item In order to analytically prove the revealed ``Exceed-Shannon'' phenomenon, we first derive an analytical finite-time capacity formula based on Mercer expansion \cite{mercer1909functions}, where we can find the connection between the capacity problem and the operator theory \cite{zhu2007operator}. To make the problem tractable, we construct a typical case in which the transmitted signal has certain statistical properties. Utilizing this construction, we obtain a closed-form capacity solution in this typical case, which leads to a rigorous proof of the ``Exceed-Shannon'' phenomenon. Inspired by the techniques in the proof, we find that the finite-time capacity is, in fact, a more general case of the Shannon limit, thus the ``Exceed-Shannon'' phenomenon of the finite-time capacity is compatible with the classical Shannon theory.
\end{itemize}

{\it Organization}: In the rest of this paper, the finite-time capacity is formulated and evaluated numerically in Section II, where the ``Exceed-Shannon'' phenomenon is first discovered. Then, in Section III, we derive a closed-form finite-time capacity formula under a typical case. Based on this formula, in Section IV, the ``Exceed-Shannon'' phenomenon is rigorously proved. Finally, conclusions are drawn in Section V.

{\it Notations}: $X(t)$ denotes a Gaussian process; $R_X(t_1, t_2)$ 
denotes the autocorrelation function; $S_X(f), S_X(\omega)$ are the power 
spectral density (PSD) of the corresponding process $X(t)$, where $\omega = 
2\pi f$; Boldface italic symbols $\bm X(t_1^n)$ denotes the column vector 
generated by taking samples of  $X(t)$ on instants $t_i, 1 \leq i \leq n$; 
Upper-case boldface letters such as ${\bf{\Phi}}$ denote matrices; 
$\mathbb{E\left[\cdot\right]}$ denotes the expectation; 
$\mathbbm{1}_{A}(\cdot)$ denotes the indicator function of the set $A$; 
$\mathcal{L}^2([0,T])$ denotes the collection of all the square-integrable 
functions on window $[0,T]$; ${\rm i}$ denotes the imaginary unit.

\section{Numerical Analysis of the Finite-Time Capacity}
	In this section, we focus on the numerical evaluation of the finite-time capacity. In Subsection~\ref{sect_2_subsect_1}, we model the transmission problem by Gaussian processes, and derive the capacity expressions within a finite-time observation window by using dense sampling and limiting methods; In Subsection~\ref{sect_2_subsect_2}, we approximate the finite-time capacity by discretized matrix-based formulas; In Subsection~\ref{sect_2_subsect_3}, we reveal the ``Exceed-Shannon'' phenomenon by numerically evaluating the finite-time capacity in a special setting of the signal autocorrelations. 

	\subsection{The Expressions for Finite-Time Capacity}\label{sect_2_subsect_1}
	The finite-time capacity is, heuristically, defined as the maximum number of bits that can be successfully transmitted within a finite-time window. Since Shannon capacity is defined on pairs of random variables, it is crucial to introduce randomness into the transmission model. Inspired by \cite{GPSampling1957BA}, we model the transmitted signal by a zero-mean stationary Gaussian stochastic process, denoted as $X(t)$, and the received signal by $Y(t):=X(t)+N(t)$. The process $N(t)$, which denotes the noise, is also a stationary Gaussian process independent of $X(t)$. The receiver is only allowed to observe the signal within finite-time window $[0,T]$, where $T>0$ is the observation window span. Our goal is to find the maximum number of bits that can be acquired within this time window. 

	To analytically express the amount of the acquired information, we first introduce $n$ sampling instants inside the time window, denoted by $(t_1, t_2 , \cdots, t_n):=t_1^n$, and then let $n \rightarrow \infty$ to approximate the finite-time capacity\footnote{In this paper, we do not explicitly distinguish between the terms ``finite-time mutual information'' and ``finite-time capacity'', since we consider communication schemes where the source autocorrelation is fixed.}. This approximation of capacity becomes more precise as the sampling instants $t_1^n$ become denser. Then by defining ${\bm X}(t_1^n)\equiv (X(t_1), X(t_2), \cdots, X(t_n))$ and ${\bm Y}(t_1^n)\equiv (Y(t_1), Y(t_2), \cdots, Y(t_n))$, the capacity on these $n$ samples can be expressed as
	\begin{equation}
		I(t_1^n) = I({\bm X}(t_1^n); {\bm Y}(t_1^n)),
		\label{eq_sample_capacity}
	\end{equation}
	and the finite-time capacity is defined as
	\begin{equation}
		I(T) = \lim_{n\rightarrow \infty}\sup_{\{t_1^n\} \subset [0,T]}{I(t_1^n)}.
		\label{eq_finite_time_capacity}
	\end{equation}

	Then, the transmission rate $C(T)$ can be defined by dividing the amount of information acquired within $[0,T]$ by the time span $T$: 
	\begin{equation}
		\begin{aligned}
		C(t_1^n)  &= I(t_1^n)/T, \\
		C(T)      & =I(T)/T.     \\
		\end{aligned}
	\end{equation}
	From these definitions, we can define the limit capacity as $C(\infty)=\lim_{T\rightarrow \infty}{C(T)}$ by letting $T\rightarrow \infty$. The quantity $C(\infty)$ characterizes the maximum average number of bits per second one can acquire from a received noisy stochastic process.

    \subsection{Discretization}\label{sect_2_subsect_2}
    Without loss of generality, we fix the sampling instants uniformly onto fractions of $T$: $t_i = (i-1)T/n, 1\leq i \leq n$. Since the random vectors ${\bm X}(t_1^n)$ and ${\bm Y}(t_1^n)$ are samples of a Gaussian process, they are both Gaussian random vectors with mean zero and covariance matrices ${\bm K}_X$ and ${\bm K}_Y$, where ${\bm K}_X, {\bm K}_Y \in \mathbb{R}^{n\times n}$ are symmetric positive-definite matrices. The entries of ${\bm K}_X$ and ${\bm K}_Y$ are determined by the autocorrelation functions of Gaussian processes $X(t)$ and $Y(t)$, denoted by $R_X(t_1, t_2)$ and $R_Y(t_1, t_2)$:
    \begin{equation}
        \begin{aligned}
            ({\bm K}_X)_{i,j} &= R_X(t_i, t_j) := \mathbb{E}\left[X(t_i)X(t_j)\right], \\
            ({\bm K}_Y)_{i,j} &= R_Y(t_i, t_j) := \mathbb{E}\left[Y(t_i)Y(t_j)\right]. \\
        \end{aligned}
    \end{equation}
    Note that $Y(t)$ is the independent sum of $X(t)$ and $N(t)$, thus the autocorrelation functions satisfy $R_Y(t_1,t_2)=R_X(t_1,t_2)+R_N(t_1,t_2)$, and similarly the covariance matrices satisfy ${\bm K}_Y={\bm K}_X+{\bm K}_N$.
    
    The mutual information $I(t_1^n)$ is defined as $I(t_1^n)=h({\bm Y}(t_1^n))-h({\bm Y}(t_1^n)|{\bm X}(t_1^n))=h({\bm Y}(t_1^n))-h({\bm N}(t_1^n))$, where $h(\cdot)$ denotes the differential entropy. For $n$-dimentional Gaussian vector ${\bm U}$ with mean 0 and covariance matrix ${\bm K}$, the differential entropy is given by
    \begin{equation}
        h({\bm U}) = \frac{1}{2} \log \left( (2\pi e)^{n}\det ({\bm K})\right).
        \label{eq_grv_diff_entropy}
    \end{equation}
	Plugging \eqref{eq_grv_diff_entropy} into the definition of $I(t_1^n)$, we obtain
    \begin{equation}
        I(t_1^n) = \frac{1}{2} \log\left(\frac{\det({\bm K}_X+{\bm K}_N)}{\det{({\bm K}_N)}} \right).
        \label{eq_I_expression}
    \end{equation}
    In \eqref{eq_I_expression}, by letting $n\rightarrow \infty$, we can find that $I(t_1^n)$ increases monotonously when $n$ doubles, because of the data processing inequality. Though without rigorous proof, we can assume with confidence that $I(t_1^n)$ is an increasing function of $n$. However, it remains unknown whether $I(t_1^n)$ tends to a finite limit. In fact, $I(t_1^n)$ can be arbitrarily large, since the signal outside the noise band is strictly unpolluted by the noise, which results in infinite SNR. Thus, the capacity will diverge to infinity. Therefore, in order to avoid capacity divergence, at least one of the following conditions should be satisfied: 
    \begin{itemize}
    	\item The noise process $N(t)$ is not band-limited.
    	\item The power spectral density of $X(t)$ is strictly contained inside the band of $N(t)$.
    \end{itemize}
	Thus, in the following numerical analysis, we choose $N(t)$ to be band-unlimited. This leads to the choice of reasonable autocorrelation functions of $X(t)$ and $N(t)$ in the following subsection.

    \subsection{Numerical Analysis}\label{sect_2_subsect_3}
        In order to study the properties of mutual information $I(t_1^n)$ as a function of $n$, we perform numerical analyses under different values of $n$ and $T$. The autocorrelation function and PSD of the signal process $X(t)$ and noise process $N(t)$ are set to the special case
        \begin{equation}
                \begin{aligned}
                    R_X(t_1, t_2) & = \mathrm{sinc}(10(t_1-t_2)),            \\
                    R_N(t_1, t_2) & = \exp(-\lvert t_1-t_2\rvert),           \\
                    S_X(f) & = 0.1\times \mathbbm{1}_{\{-5\leq f \leq 5\}},  \\
                    S_N(f) & = \frac{2}{1+(2\pi f)^2},                       \\
                \end{aligned}
            \label{eq_correlation_examples_1}
        \end{equation}
        where $\mathrm{sinc}(x):=\sin(\pi x)/(\pi x)$. Note that the PSD of the transmitted process $S_X(f)$ is strictly band-limited, while the PSD of the noise process is not. In fact, the noise PSD is carefully selected to ensure the received noise has finite power on each instant $t_i$, allowing the execution of numerical computations. A finitely powered process PSD must be colored, in contrast to additive white Gaussian noise (AWGN) with white PSD and infinite power. That is the reason why we choose $S_N(f)$ to be the form of \eqref{eq_correlation_examples_1}.
		
		In order to compare the finite-time capacity with the classical Shannon capacity, we have to calculate the Shannon capacity with colored noise spectrum $S_N(f)$, which is a generalized version of the well-known formula $C=W\log\left(1+S/N\right)$. The Shannon capacity $C_{\rm sh}$ of colored noise PSD \cite{Cover1999ElementsInfTheory}, measured in ${\rm nat/s}$, is expressed as 
        \begin{equation}
            C_{\rm sh} := \frac{1}{2} 
            \int_{-\infty}^{+\infty}{\log\left(1+\frac{S_X(f)}{S_N(f)}\right)\mathrm{d}
             f} \quad [{\rm nat/s}].
            \label{eq_shannon_formula}
        \end{equation}
        Then, plugging \eqref{eq_correlation_examples_1} into \eqref{eq_shannon_formula} yields the numerical result for $C_{\rm sh}$.
        
\begin{figure}[htbp]
            \centering
            \myincludegraphics{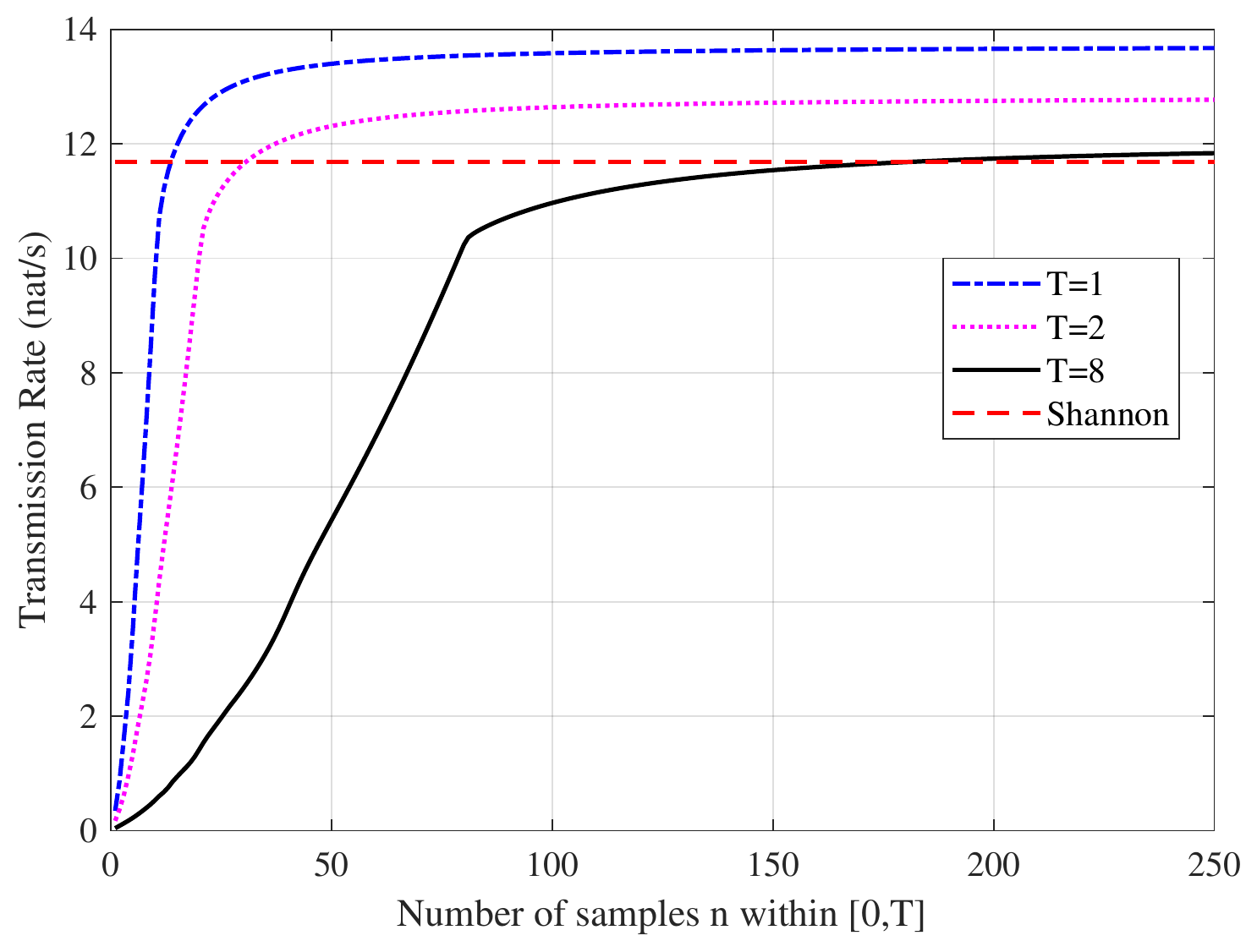}
            \caption{A first glance to the ``Exceed-Shannon'' phenomenon, where the red dashed horizontal line is the Shannon capacity, and the $T=1, 2, 8$ curves illustrate the dependence of $C(t_1^n)$ on $n$.}
            \label{fig_sp_finite_time_rate_wrt_n}
        \end{figure}

        In the numerical analysis, we calculate the finite-time transmission rate $C(T)$ and Shannon capacity against the number of samples $n$ within the observation window $[0,T]$. The numerical results are collected in Fig.~\ref{fig_sp_finite_time_rate_wrt_n}. It is shown that $I(t_1^n)$ is an increasing function of $n$, and for fixed values of $T$, the approximated finite-time capacity $I(t_1^n)$ tends to a finite limit under the correlation assumptions given by \eqref{eq_correlation_examples_1}. The most amazing observation is that, we can obtain more information within finite-time window $[0,T]$ than the prediction $T C_{\rm sh}$ given by the Shannon capacity \eqref{eq_shannon_formula}. We call this phenomenon the ``Exceed-Shannon'' phenomenon.

    	To analytically verify the existence of the Exceed-Shannon phenomenon, we are going to introduce some mathematical tools in the following Section III, and finally give an analytical proof in Section IV.
        
\section{A Closed-Form Finite-Time Capacity Formula}
	In this section, we first introduce the Mercer expansion in Subsection~\ref{sect_3_subsect_1} as a basic tool for our analysis. Then we derive the series representation of the finite-time capacity, and the corresponding power constraint in Subsection~\ref{sect_3_subsect_2}, under the assumption of AWGN noise. The power constraint shows that the finite-time capacity is upper-bounded, thus the series expansion of the finite-time capacity converges absolutely.
	
    \subsection{The Mercer Expansion}\label{sect_3_subsect_1}
    Motivated by the discovery of the Exceed-Shannon phenomenon, we go further into the underlying mechanism behind this fact. Since the calculation of \eqref{eq_I_expression} depends on the evaluation of determinants that are determined by autocorrelation functions of Gaussian processes, it is possible to obtain $I(t_1^n)$ and $I(T)$ directly from the autocorrelation functions. In fact, if we know the Mercer expansion \cite{mercer1909functions} of the autocorrelation function $R_X(t_1, t_2)$ on window $[0,T]$, then we can calculate $h({\bm X}(t_1^n))$ more easily \cite{MercerExpansion1955Barrett}. In the following discussion, we assume the Mercer expansion of the source autocorrelation function $R_X(t_1, t_2)$ to be in the following form
    \begin{equation}
        \lambda_k \phi_k(t_1) = \int_{0}^{T}{R_X(t_1, t_2) \phi_k(t_2) \mathrm{d}t_2}; k > 0, k\in \mathbb{N}.
        \label{eq_mercer_expansion}
    \end{equation}
    Due to the positive-definite property of the integral kernel $R_X(t_1, t_2)$, the eigenvalues are strictly positive: $\lambda_k > 0$, and the eigenfunctions form an orthonormal set:
    \begin{equation}
        \int_{0}^{T}{\phi_i(t)\phi_j(t)\mathrm{d}t} = \delta_{ij}.
        \label{eq_orthonormal}
    \end{equation}
        
    The Mercer's theorem \cite{mercer1909functions} ensures the existence and uniqueness of the eigenpairs $(\lambda_k, \phi_k(t))_{k=1}^{\infty}$, and furthermore, the kernel itself can be expanded under the eigenfunctions. The convergence is absolute and uniform:
    \begin{equation}
        R_X(t_1, t_2) = \sum_{k=1}^{+\infty}{\lambda_k \phi_k(t_1)\phi_k(t_2)}.
        \label{eq_mercer_kernel_expansion}
    \end{equation}
	The Mercer expansion enables us to analytically express an autocorrelation function on a finite-time interval $[0,T]$, since the autocorrelation function can be naturally treated as a positive-definite integral kernel.
    
    \subsection{Finite-Time Capacity Formula} \label{sect_3_subsect_2}
    Based on Mercer expansion, we can obtain a closed-form formula in the following {\bf Theorem \ref{th_1}}.
        \begin{theorem}[Source expansion, AWGN noise]\label{th_1}
            Suppose the information source, modeled by the transmitted process $X(t)$, has autocorrelation function $R_X(t_1, t_2)$. An AWGN noise of PSD $n_0/2$ is imposed onto $X(t)$, resulting in the received process $Y(t)$. The Mercer expansion of $R_X(t_1, t_2)$ on $[0,T]$ is given by \eqref{eq_mercer_expansion}, satisfying \eqref{eq_orthonormal}. Then the finite-time mutual information $I(T)$ within the observation window $[0,T]$ between the processes $X(t)$ and $Y(t)$ can be expressed as
            \begin{equation}
                I(T)=\frac{1}{2} \sum_{k=1}^{+\infty}{\log\left(1+\frac{\lambda_k}{n_0/2}\right)}.
                \label{eq_th_1}
            \end{equation}
        \end{theorem}
    	\begin{IEEEproof}
    		See Appendix A.
    	\end{IEEEproof}

        From {\bf Theorem \ref{th_1}}, we can conclude that the finite-time capacity of AWGN channel is uniquely determined by the Mercer spectra $\lambda_k$ of $R_X(t_1, t_2)$ within $[0,T]$. However, it remains unknown whether the series representation \eqref{eq_th_1} converges. In fact, the convergence is closely related to the signal power. In Fourier transform, the power in the time domain is equal to the power in the frequency domain, which is known as the Parseval's theorem\cite{Kelkar1983parsevaltheorem}. Like the Fourier transform, the transform defined by the orthonormal basis $\{\phi_k(t)\}_{k=1}^{\infty}$ also satisfy the Parseval's theorem. This observation leads to a theoretic verification of power conservation in the view of $\lambda_k$, which is stated in the following {\bf Lemma \ref{lemma_1}}.
            \begin{lemma}[Operator Trace Coincide with Power Constraint]
                Given stationary Gaussian process $X(t)$ with mean zero and autocorrelation $R_X(t_1, t_2)$. The Mercer expansion of $R_X(t_1, t_2)$ on $[0,T]$ is given by \eqref{eq_mercer_expansion}, satisfying \eqref{eq_orthonormal}. The Mercer operator $M(\cdot): \mathcal{L}^2([0,T]) \rightarrow \mathcal{L}^2([0,T])$ is defined by the integral $(M\phi)(s) = \int_{0}^{T}{R_X(s ,\tau)\phi(\tau)\mathrm{d}\tau}$. Then the sum of all the eigenvalues $\lambda_k$ of operator $M$ is equal to the signal energy $PT$ within $[0,T]$:
                \begin{equation}
                    \mathrm{tr}(M) := \sum_{k=1}^{+\infty}{\lambda_k} = PT,
                \end{equation}
                where $P=R_X(0,0)$.
                \label{lemma_1}
            \end{lemma}
        	\begin{IEEEproof}
        		See Appendix B.
        	\end{IEEEproof}

			\begin{remark}\label{remark_1}
				The convergence of the finite-time capacity series~\eqref{eq_th_1} is ensured by {\bf Lemma \ref{lemma_1}}. In fact, from the above {\bf Lemma \ref{lemma_1}}, we can conclude that the sum of $\lambda_k$ is finite when $T$ is finite. It can be immediately derived that $I(T)<\infty$, since $\log(1+x) \leq x$. Furthermore, note that the sum of $\lambda_k$ is finite even for non-stationary processes (i.e., the power at time $t$: $P(t):=\mathbb{E}\left[X^2(t)\right]$ is not always a constant $P:=R_X(0,0)$), as long as $P(t)<\infty$ holds for any $0<t<T$. Then the conclusion $I(T)<\infty$ holds even for non-stationary processes.
            \end{remark}

            \begin{remark}
                The finite-time capacity formula~\eqref{eq_th_1} is closely related to the operator theory \cite{zhu2007operator} in functional analysis. The sum of all the eigenvalues $\lambda_k$ is called the operator trace in linear operator theory. As is mentioned in {\bf Lemma \ref{lemma_1}}, the autocorrelation function $R_X(t_1, t_2)$ can be treated as a linear operator $M$ on $\mathcal{L}^2([0,T])$. Furthermore, this operator belongs to the trace class \cite{brislawn1988kernels} if and only if $\int_{0}^{T}{R_X(t,t)\mathrm{d}t}<\infty$. Note that this condition is automatically satisfied if $X(t)$ is a Gaussian process, since Gaussian random variables always have finite variances.
			\end{remark}
            
			The Mercer spectra enables us to explicitly calculate the finite-time capacity, and furthermore, prove the ``Exceed-Shannon'' phenomenon. This will be demonstrated in the next section. 

\section{Proof of the Existence of Exceed-Shannon Phenomenon}
    In this section, we first give two different proofs of the existence of the Exceed-Shannon phenomenon, both in a typical case. Then we discuss the achievability of the finite-time capacity, and the compatibility with Shannon-Hartley Theorem.
    \subsection{Closed-Form Capacity in A Typical Case} \label{sect_4_subsect_1}
        In order to show the existence of Exceed-Shannon phenomenon, we only need to show that the finite-time capacity is greater than Shannon capacity in a typical case. Let us consider a finite-time communication scheme with a finitely-powered stationary transmitted signal autocorrelation\footnote{The signal autocorrelation $R_X(\tau)$ is often observed in many scenarios, such as passing a signal with white spectrum through an RC lowpass filter.}, which is specified as 
        \begin{equation}
            R_X(t_1, t_2) = R_X(\tau) = P \exp(-\alpha \lvert \tau\rvert),
            \label{eq_theoretical_verification_scheme}
        \end{equation}
        where $\tau = t_1 - t_2$, in AWGN channel\footnote{In this theoretical proof of ``Exceed-Shannon'' phenomenon, we assume the noise to be AWGN, to simplify the analytical computations. Gaussian processes of white spectrum are ``immoral'', thus they can neither be power-limited, nor they can be directly sampled and numerically represented in computers.} with noise PSD being $n_0/2$. The power of signal $X(t)$ is $P=R_X(0)$. According to {\bf Lemma~\ref{lemma_1}}, the trace of the corresponding Mercer operator $M(\cdot)$ is finite. Then the finite-time capacity given by {\bf Theorem~\ref{th_1}} is also finite, as is shown in {\it Remark~\ref{remark_1}}. Finding the Mercer expansion is equivalent to finding the eigenpairs $(\lambda_k, \phi_k(t))_{k=1}^{\infty}$. The eigenpairs are determined by the following characteristic integral equation \cite{Cai2020Eigenvalue}:
        \begin{equation}
            \lambda_k \phi_k(s) = \int_{0}^{s}{P e^{-\alpha (s-t)} \phi_k(t) \mathrm{d}t}+\int_{s}^{T}{P e^{-\alpha (t-s)} \phi_k(t) \mathrm{d}t}.
            \label{eq_integral_eqn}
        \end{equation}
        Differentiating both sides of \eqref{eq_integral_eqn} twice with respect to $s$ yields the boundary conditions and the differential equation that $\lambda_k, \phi_k$ must satisfy:
        \begin{equation}
            \begin{aligned}
                \lambda_k \phi_{k}''(s) & = (\alpha^2 \lambda_k - 2\alpha P)\phi_k(s), 0<s<T, \\
                \phi_k'(0) & = \alpha \phi_k(0), \\
                \phi_k'(T) & = -\alpha \phi_k(T). \\
            \end{aligned}
        \end{equation}
        Let $\omega_k>0$ denote the resonant frequency of the above harmonic oscillator differential equation, then $\lambda_k = 2\alpha P / (\alpha^2 + \omega_k^2)$, and $\omega_k$ must satisfy the above two boundary constraints. Let $\phi_k(t)=A_k \cos(\omega_k t) + B_k \sin(\omega_k t)$ be the sinusoidal form of the eigenfunction. Using the boundary conditions we obtain
		\begin{equation}
            \begin{aligned}
                & B_k \omega_k  = \alpha A_k, \\
                & B_k \omega_k \cos(\omega_k T)-A_k \omega_k \sin(\omega_k T) \\
				& =-\alpha \left( A_k \cos(\omega_k T) + B_k \sin(\omega_k T)\right).  \\
            \end{aligned}
            \label{eq_determine_omega_1}
        \end{equation}
        To ensure the existence of solution to the homogeneous linear equations \eqref{eq_determine_omega_1} with unknowns $A_k, B_k$, the determinant must be zero. Exploiting this condition, we find the equation that $\omega_k$ must satisfy:
        \begin{equation}
            \tan(\omega_k T) = \frac{2\omega_k \alpha}{\omega_k^2 -\alpha^2}.
            \label{eq_determine_omega_2}
        \end{equation}
        \begin{figure}[!t]
            \centering
            \myincludegraphics{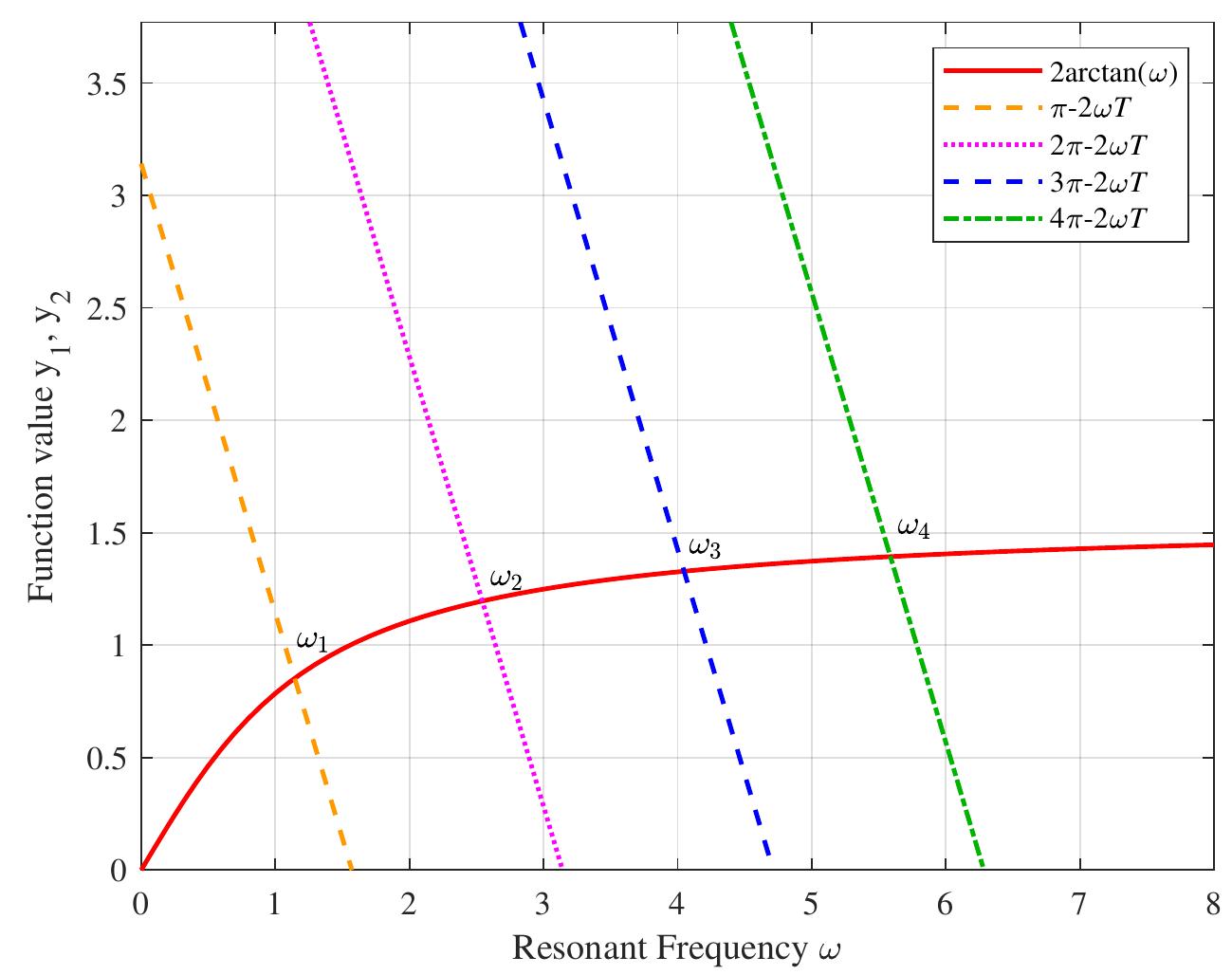}
            \caption{Images of $2\arctan(\omega/\alpha)$ and $k\pi - \omega T$ with $T=2, \alpha=1$ and $k=1,2,3,4$. The desired resonant frequencies $\omega_k$ can be read from the horizontal coordinates of the intersection points of the curve $2\arctan(\omega/\alpha)$ and the parallel lines. }
            \label{fig_eigen_determine_curves}
        \end{figure}
        By introducing an auxillary variable $\theta_k = \arctan(\omega_k/\alpha) \in [0, \pi/2]$, equation \eqref{eq_determine_omega_2} can be simplified as $\tan(\omega_k T) = -\tan(2\theta_k)$, i.e., there exists a positive integer $m$ such that $2\arctan(\omega_k / \alpha) = m\pi - \omega_k T$. The integer $m$ can be chosen to be equal to $k$. From the function images of $2\arctan(\omega/\alpha)$ and $m\pi - \omega T$ (Fig.~\ref{fig_eigen_determine_curves}), we can determine $\omega_k$, and then $\lambda_k$. To sum up, the solution to the characteristic equation \eqref{eq_integral_eqn} are collected into \eqref{eq_eigenpairs} as follows.
        \begin{equation}
            \begin{aligned}
                2\arctan(\omega_k /\alpha) & = k\pi - \omega_k T,       \\
                \lambda_k & = \frac{2\alpha P}{\alpha^2 + \omega_k^2},   \\
                \phi_k(t) & = \frac{1}{Z_k} \left(\omega_k \cos(\omega_k t) + \alpha \sin(\omega_k t)\right), \\
            \end{aligned}
            \label{eq_eigenpairs}
        \end{equation}
		where $Z_k$ denotes the normalization constants of $\phi_k(t)$ on $[0,T]$ to ensure orthonormality.

        Equation \eqref{eq_eigenpairs} gives all eigenpairs $(\lambda_k, \phi_k)_{k=1}^{\infty}$, from which we can calculate $I(T)$ by applying {\bf Theorem~\ref{th_1}}. As for the Shannon capacity $C_{\rm sh}$, by applying \eqref{eq_shannon_formula} and evaluating the integral with\cite{integralTable2014}, we can obtain
        \begin{equation}
            \begin{aligned}
                C_{\rm sh} & = 
                \frac{1}{4\pi}\int_{-\infty}^{+\infty}{\log\left(1+\frac{S_X(\omega)}{n_0/2}
                 \right)\mathrm{d}\omega} \\
                & = 
                \frac{1}{4\pi}\int_{-\infty}^{+\infty}{\log\left(1+\frac{\frac{2P\alpha}{\alpha^2+\omega^2}}{n_0/2}
                 \right)\mathrm{d}\omega} \\
                & \overset{(a)}{=} \frac{1}{2} \left(\sqrt{\alpha^2+\frac{4P\alpha}{n_0}} - \alpha\right),
            \end{aligned}
            \label{eq_Csh_example2}
        \end{equation}
    	where the evaluation of the improper integral $(a)$ is given in Appendix E.

        After all the preparation works above, we can rigorously prove that $C(T) > C_{\rm sh}$ under the typical case of \eqref{eq_theoretical_verification_scheme}, as long as the transmission power $P$ is smaller than a constant $\delta$. The following {\bf Theorem~\ref{th_2}} proves this result.

        \begin{theorem}[Existence of Exceed-Shannon phenomenon in a typical case] \label{th_2}
            Suppose $X(t)$ and $Y(t)$ are specified according to \eqref{eq_theoretical_verification_scheme}. The eigenpairs are determined by \eqref{eq_eigenpairs}. Then, for any fixed positive values of $T, n_0$ and $\alpha$, there exists a positive number $\delta$ such that the Exceed-Shannon inequality
            \begin{equation}
                C(T):=\frac{1}{2T}\sum_{k=1}^{+\infty}\log\left(1+\frac{\lambda_k}{n_0/2} \right) > C_{\rm sh}
                \label{eq_exceed_shannon_special}
            \end{equation}
            holds strictly for arbitrary $0<P<\delta$.
        \end{theorem}
        \begin{IEEEproof}
            See Appendix C.
        \end{IEEEproof}

        \begin{figure}[t]
            \centering
            \myincludegraphics{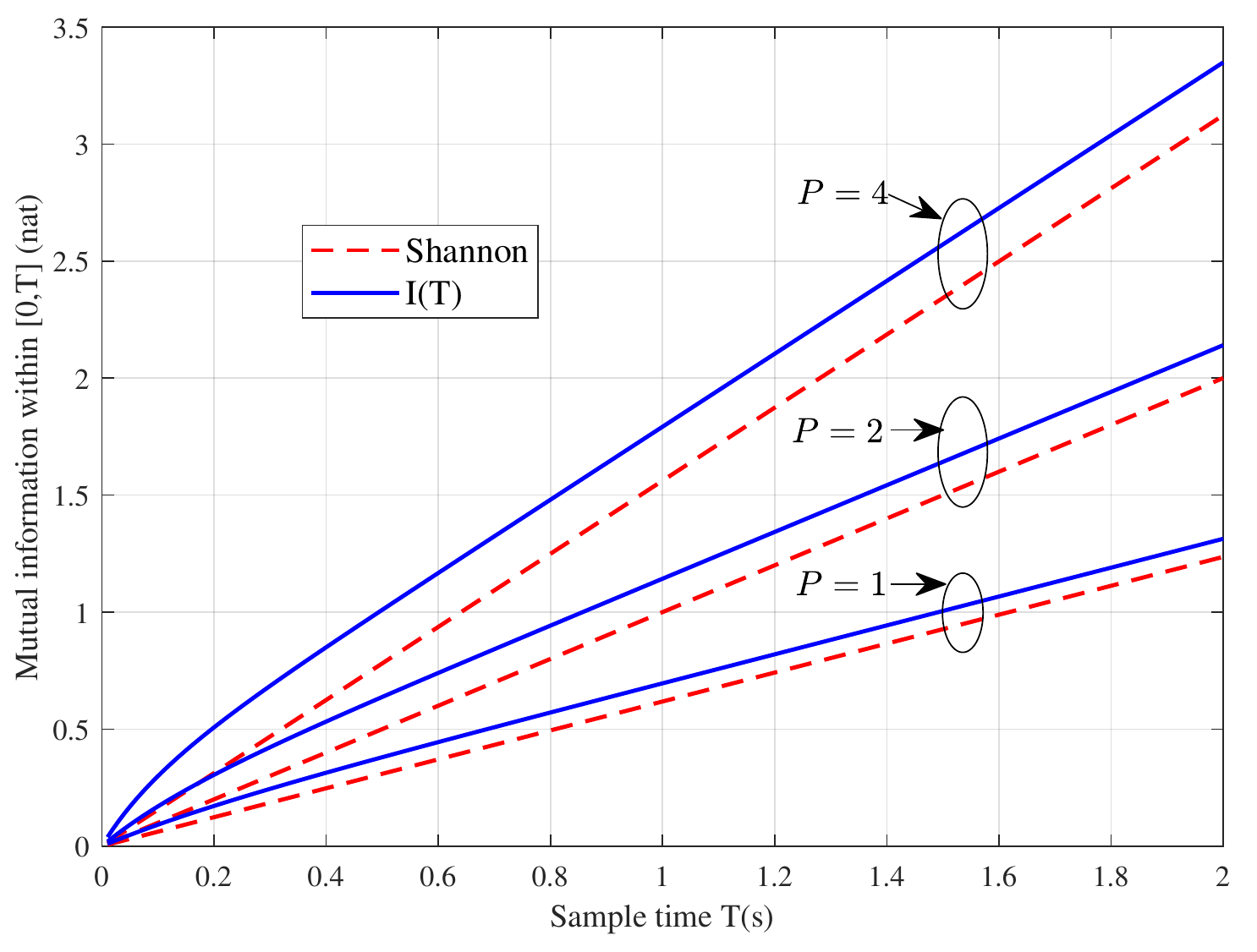}
            \caption{Theoretical verification of the Exceed-Shannon effect. The blue lines represent the finite-time mutual information $I(T)$. The red lines are the Shannon capacities $TC_{\rm sh}$ calculated from \eqref{eq_shannon_formula}. All curves are evaluated under hypothesis \eqref{eq_theoretical_verification_scheme}, where $P=1,2,4$, $n_0 = 1$ and $\alpha=1$.}
            \label{fig_theoretical_exceed_shannon}
        \end{figure}

        To verify the above theoretical analysis, numerical experiments on $I(T)$ are conducted based on evaluations of \eqref{eq_eigenpairs} and \eqref{eq_Csh_example2}. As is shown in Fig.~\ref{fig_theoretical_exceed_shannon}, it seems that we can always harness more mutual information in a finite-time observation window, compared with the Shannon capacity. Though seems impossible, this fact is somehow unsurprising because the observations ${\bm Y}(t_1^N)$ inside the finite-time window $[0,T]$ can always eliminate some extra uncertainty outside the window due to the autocorrelation of $X(t)$. Different from the finite-time capacity, the Shannon capacity describes the circumstance of $T\rightarrow \infty$, where the fringe effect near $t=0$ and $t=T$ becomes negligible compared with the prolonged window period. Thus, the Shannon capacity does not take into consideration the small extra information on the fringe, causing an underestimation of the capacity. Fig.~\ref{fig_theoretical_exceed_shannon} also shows that, the extra capacity $\Delta I := I(T) - TC_{\rm sh}$ between the finite-time result and Shannon capacity tends to a constant as $T\rightarrow \infty$. As is discussed above, the difference may come from the additional elimination of uncertainty at the fringe of the window. This asymptotically constant difference results in asymptotic linearity of the finite-time mutual information $I(T)$ as a function of $T$.
        
        Apart from the above discussion, there is an extra interesting observation in Fig.~\ref{fig_theoretical_exceed_shannon}, which leads to another rigourous proof of the Exceed-Shannon phenomenon. If we investigate the slope of curves $I(T)$ at the origin, i.e., the ``instant transmission rate'' at the origin, we find that $C(0^{+}) > C_{\rm sh}$. This observation is confirmed by the following theorem:
        
        \begin{theorem}[Instant Finite-Time Rate Exceeds Shannon]\label{th_3}
            Suppose $X(t)$ and $Y(t)$ are specified according to \eqref{eq_theoretical_verification_scheme}. The eigenpairs are given by \eqref{eq_eigenpairs}. Then the instantaneous finite-time information transmission rate $C(0^{+})$ is given by:
            \begin{equation}
            	C(0^{+}) = \frac{\partial I(T)}{\partial T} |_{T=0^+} = \frac{P}{n_0}
            	\label{eq_instant_rate_at_origin}
            \end{equation}
        \end{theorem}
    	\begin{IEEEproof}
    		See Appendix D.
    	\end{IEEEproof}
    
    	From the conclusion of {\bf Theorem~\ref{th_3}} and \eqref{eq_Csh_example2}, we can reduce the Exceed-Shannon inequality $C(0^{+}) > C_{\rm sh}$ at $T=0^{+}$ to the following inequality:
    	\begin{equation}
    		\frac{P}{n_0} > \frac{1}{2} \left(\sqrt{\alpha^2+\frac{4P\alpha}{n_0}} - \alpha\right),
    		\label{eq_exceed_shannon_ineq_at_origin}
    	\end{equation}
        which can be directly verified by simple term-shifting and squaring on both sides. This inequality implies that the average transmission rate in the finite-time regime is strictly larger than the Shannon capacity around the origin $T=0$. 
        
        \begin{remark}
        	Both {\bf Theorem~\ref{th_2}} and {\bf Theorem~\ref{th_3}} prove the existence of Exceed-Shannon phenomenon rigorously. However, their differences are:
        	\begin{itemize}
        		\item {\bf Theorem~\ref{th_2}} proves that the Exceed-Shannon inequality holds in a small power range $0<P<\delta$, but {\bf Theorem~\ref{th_3}} is true for arbitrary power $P$.
        		\item Since {\bf Theorem~\ref{th_3}} only proves $C(0^+)>C_{\rm sh}$, it does not ensure the Exceed-Shannon inequality when $T$ becomes larger. By contrast, {\bf Theorem~\ref{th_2}} states that the the Exceed-Shannon phenomenon exists for arbitrary $T>0$.
        	\end{itemize}
        \end{remark}

		\begin{remark}
			In fact, {\bf Theorem~\ref{th_2}} and {\bf Theorem~\ref{th_3}} characterize the Exceed-Shannon phenomenon of the finite-time capacity from two aspects.  One aspect is the observation time $T$, and the other is the transmit power $P$. Combining these two proofs of the theorems may result in a universal proof that is independent of the choice of parameters $T$ and $P$, which requires further study.
		\end{remark}

\begin{figure}[t]
			\centering
			\myincludegraphics{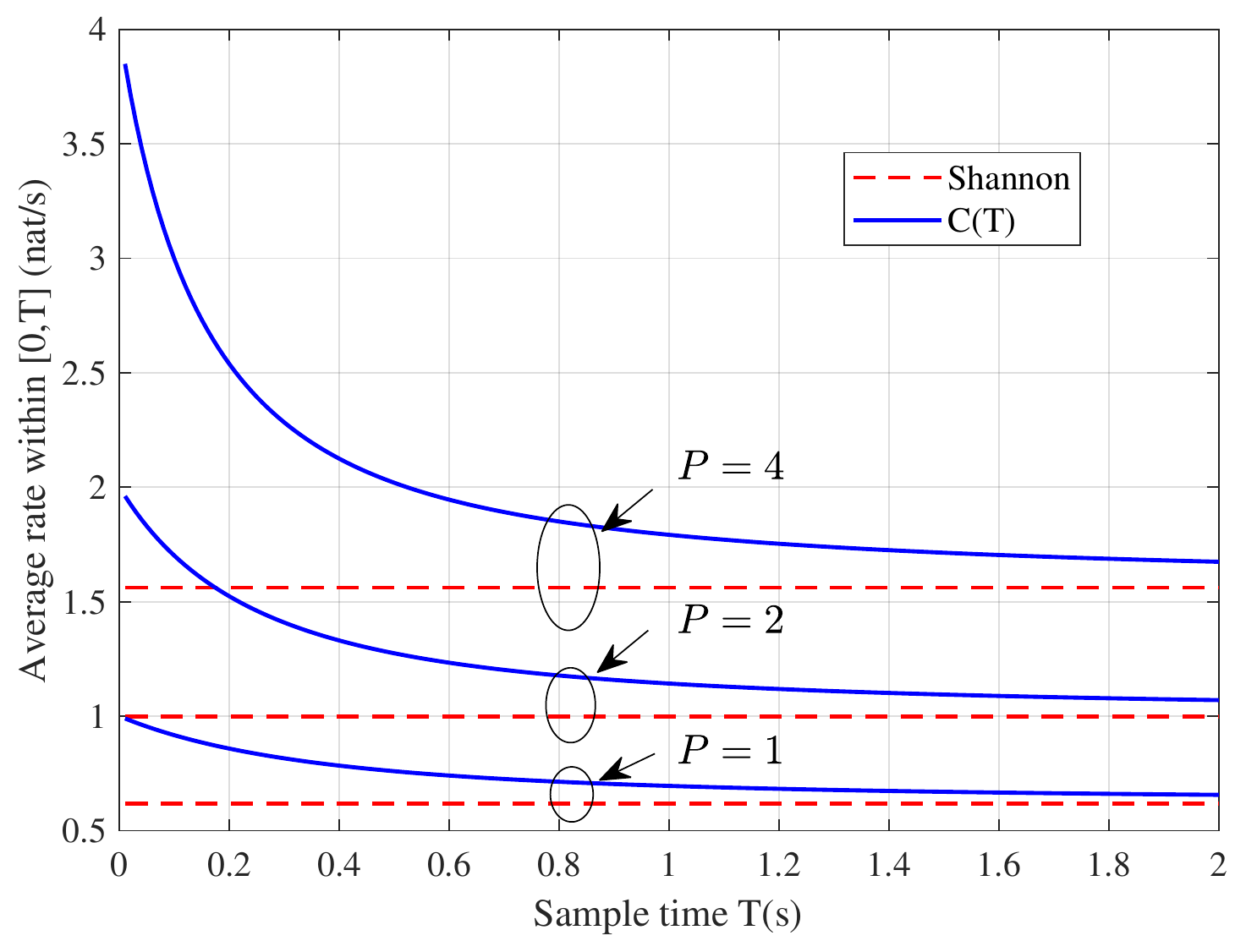}
			\caption{Comparison between the finite-time capacity $C(T)$ and Shannon capacity, when $\alpha = 1$.}
			\label{fig_CT_alpha_1}
			\centering
			\myincludegraphics{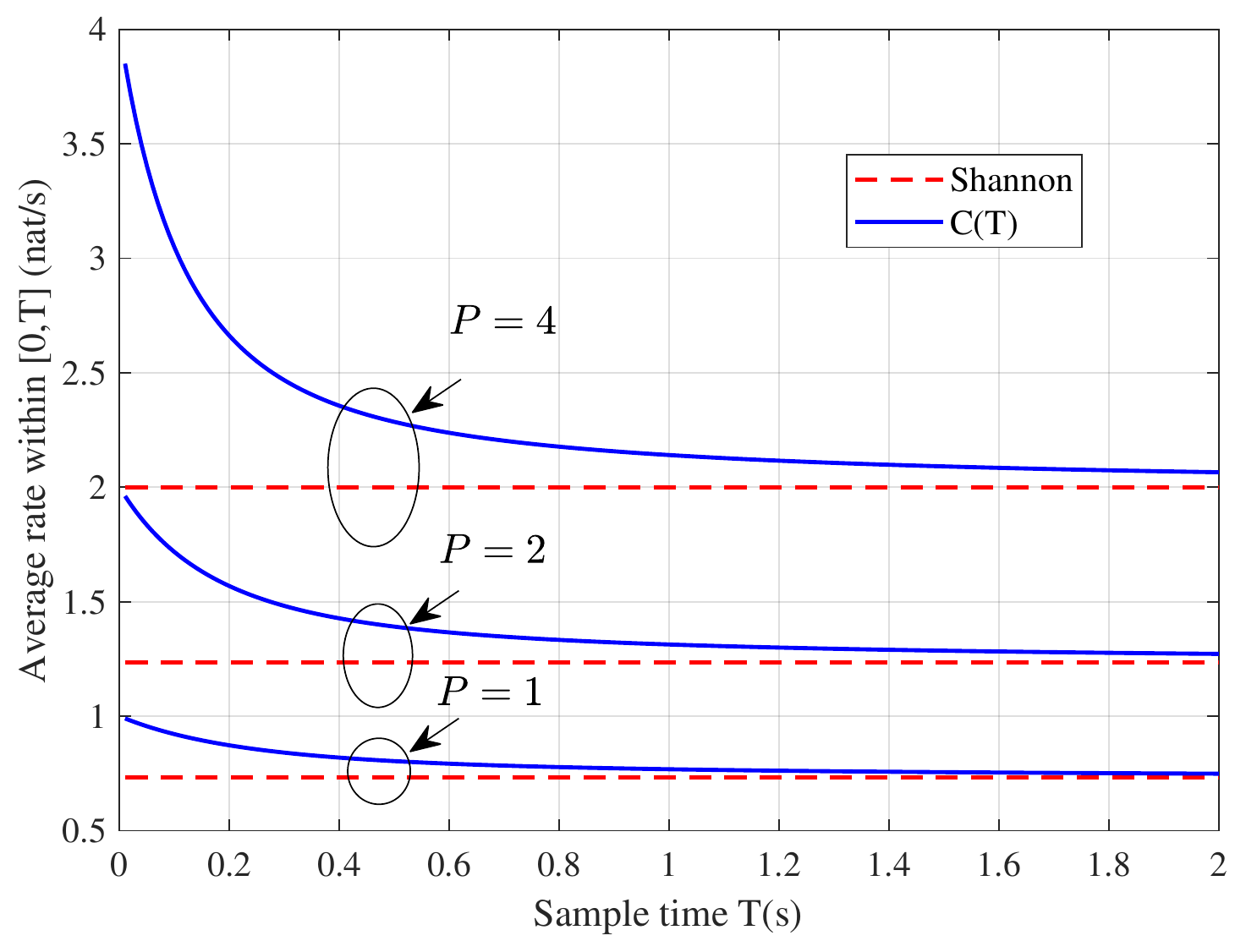}
			\caption{Comparison between the finite-time capacity $C(T)$ and Shannon capacity, when $\alpha = 2$.}
			\label{fig_CT_alpha_2}
		\end{figure}

		The conclusion of {\bf Theorem~\ref{th_3}} is verified numerically in Fig.~\ref{fig_CT_alpha_1} and Fig.~\ref{fig_CT_alpha_2}. The blue solid lines, representing the finite-time capacity $C(T)$, are above the red dashed lines representing the Shannon capacity, which demonstrates the Exceed-Shannon phenomenon. The $C(T)$ curves all start at $P/n_0$ when $T=0^+$, which coincides with the conclusion of {\bf Theorem~\ref{th_3}}. It can also be observed from the two figures that, for fixed values of $P,n_0$, as $\alpha$ increases, the transmitted signal $X(t)$ tends to be less correlated, thus being more informative. The transmission rate is then improved. This insight also comes from the change of PSD $S_X(f)$. As $\alpha$ increases, the PSD becomes flatter, i.e., a wider range of bandwidth is occupied, and thus the rate increases accordingly.   
		
    \subsection{Further Discussions on the Exceed-Shannon Phenomenon}\label{sect_4_subsect_2}
        {\bf Achievability of the finite-time capacity.} It is known that for any band-limited stationary Gaussian process $X(t)$ with PSD $S_X(f)$, one can generate signals whose PSD is exactly $S_X(f)$ by first generating a sequence $X(nT_s)$ of sufficiently high sampling rate, and then passing the generated sequence through a shaping filter. Since the transmitted signal $X(t)$ and its generating sequence $X(nT_s)$ determine each other uniquely if $X(t)$ is strictly band-limited, $X(t)$ and $X(nT_s)$ can be treated to contain the same amount of information. Then we can conclude that, after observing the noisy received process $Y(t), t\in [0,T]$, because of the definition of the finite-time capacity $I(T)$, the amount of uncertainty of the underlying transmitted sequence $X(nT_s)$ can be reduced by exactly $I(T)$ nats. That is to say, we link the finite-time capacity with a sequence-to-sequence capacity. Thus, the finite-time mutual information $I(T)$ is achievable by standard capacity-achieving techniques such as random coding \cite{Shannon}, as long as the sampling instants are dense enough.

        {\bf Compatibility with the Shannon-Hartley theorem.} Though the Exceed-Shannon effect does imply an average data transmission rate within a finite-time window higher than that predicted by Shannon, in fact, it is still impossible to construct a long-time stable data transmission scheme above the Shannon capacity by leveraging this effect. So the Exceed-Shannon phenomenon does not contradict the Shannon-Hartley Theorem. Placing additional observation windows cannot increase the average information rate, because the received process $Y(t)$ observed by the subsequent additional windows has already been implicitly altered by the previous observation. The posterior process $Y(t)|{\bm Y}(t_1^N)$ does not carry as much information as the original one, thus causing a rate reduction in the later windows. It is expected that, the average transmission rate would ultimately decrease to the Shannon capacity as the total observation time tends to infinity (i.e., $C(\infty)=C_{\rm sh}$), and the analytical proof is still worth investigation in future works.

\section{Conclusions}    
    In this paper, we provided rigorous proofs of the existence of the ``Exceed-Shannon'' phenomenon under typical autocorrelation settings of the transmitted signal process and the noise process. Our discovery of the ``Exceed-Shannon'' phenomenon revealed a possible new direction of research in information theory, as it provided a generalization of Shannon's renowned formula $C=W\log(1+S/N)$ to the practical finite-time communications. It shows the possibility that we can communicate at a higher-than-Shannon rate in a short time. Since the finite-time capacity is a more precise estimation of the ultimate capacity limit, the optimization target may shift from the Shannon capacity to the finite-time capacity in the design of practical communication systems. Thus, it has guiding significance for the performance improvement of modern communication systems. In future works, general proofs of $C(T)>C_{\rm sh}$, independent of the concrete autocorrelation settings, still require further investigation. Moreover, we need to answer the question of how to exploit this Exceed-Shannon phenomenon to improve the communication rate. In addition, although we have discovered numerically that the finite-time capacity agrees with the Shannon capacity when $T\rightarrow \infty$, an analytical proof of this result is required in the future.


\section*{Appendix A \\ Proof Of Theorem \ref{th_1}}
	Define $n$-by-$n$ matrix ${\bm \Phi}_n$ as $${\bm \Phi}_n = [\phi_1(t_1^n), \phi_2(t_1^n), \cdots, \phi_n(t_1^n)],$$ where $t_i = (i-1)T/n, 1\le i \le n$. According to the definition of this matrix, the following relation holds:
	\begin{equation}
		\begin{aligned}
			\left(\frac{T}{n}{\bm \Phi}_{n}^{\mathrm{T}}{\bm \Phi}_n\right)_{ij} & = \frac{T}{n}\sum_{k=1}^{n}{\phi_i(t_k)\phi_j(t_k)} \\
			& \rightarrow \int_{0}^{T}{\phi_i(t)\phi_j(t)\mathrm{d}t} = \delta_{ij}. \\
		\end{aligned}
		\label{eq_phi_n_orthogonal}
	\end{equation}
	This implies that the matrix ${\bm \Phi}_n$ satisfies the property of asymptotic orthogonality:
	\begin{equation}
		\Vert \frac{T}{n}{\bm \Phi}_{n}^{\mathrm{T}}{\bm \Phi}_n - I_n \Vert_2 \rightarrow 0,
	\end{equation}
	and the matrix ${\bm \Phi}_n$ can asymptotically diagonalize ${\bm K}_X=\left(R_X(t_i, t_j)\right)_{i,j=1}^{n}$ because of the eigenvalue property \eqref{eq_mercer_expansion}:
	\begin{equation}
		\begin{aligned}
			\mathbb{E}\left[\frac{T^2}{n^2} {\bm \Phi}_{n}^{\mathrm{T}} {\bm X}(t_1^n) {\bm X}^{\mathrm{T}}(t_1^n) {\bm \Phi}_{n} \right] & = \frac{T^2}{n^2} {\bm \Phi}_{n}^{\mathrm{T}} {\bm K}_X {\bm \Phi}_{n} \\ & \rightarrow \mathrm{diag}(\lambda_1, \lambda_2, \cdots, \lambda_n).
		\end{aligned}
		\label{eq_diagonalize_Kx}
	\end{equation}
	
	Next, we investigate the noise realizations on sampling instants $t_i$: For AWGN noise, the instantaneous power is $\infty$, i.e., $\mathbb{E}[N(t_j)^2]=\infty$, so it is necessary to assume that the AWGN noise is sampled after passing a rectangular-shaped impulse response filter $\xi(t)$ with pulse width $T/n$ and gain $n/T$. This assumption is reasonable, since the filter $\xi(t)$ tends to an ideal sampler $\delta(t)$ as $n\to  \infty$. Under this hypothesis, the noise variance for each sample can be calculated as
	\begin{equation}
		\begin{aligned}
			& \mathbb{E}\left[\left(\int_{0}^{T/n}{N(t)\xi\left(\frac{T}{n}-t\right)\mathrm{d}t}\right)^2\right] \\ 
			& = \iint_{0}^{T/n}{\mathrm{d}t_1\mathrm{d}t_2}{\mathbb{E}\left[N(t_1)N(t_2)\right]  \xi\left(t_1\right)\xi\left(t_2\right)}  \\ 
			& \overset{(a)}{=} \iint_{0}^{T/n}{\mathrm{d}t_1\mathrm{d}t_2} {\frac{n_0}{2} \delta(t_1-t_2) \xi\left(t_1\right)\xi\left(t_2\right)}  \\ 
			& = \frac{n_0}{2} \int_{0}^{T/n}{\xi^{2}(t) \mathrm{d}t} \\ 
			& = \frac{n_0}{2}\frac{n}{T}. \\
		\end{aligned}
	\end{equation}
	Note that the equality $(a)$ holds since the noise autocorrelation is $\frac{n_0}{2} \delta(t_1-t_2)$. In this way, the mutual information within window $[0,T]$ can be calculated as
	\begin{equation}
		\begin{aligned}
			I(T)  = & \frac{1}{2} \lim_{n\rightarrow \infty} \log\left(\frac{\det({\bm K}_X + \frac{nn_0}{2T}{\bm I}_n)}{\det\left(\frac{nn_0}{2T}{\bm I}_n\right)}\right) \\
			= & \frac{1}{2} \lim_{n\rightarrow \infty}\log \det\left({\bm I}_n + \frac{2T}{nn_0} {\bm K}_X \right) \\
			\overset{(b)}{=} & \frac{1}{2} \lim_{n\rightarrow \infty} \log \det \left(\frac{T}{n}{\bm \Phi}_{n}^{\mathrm{T}}{\bm \Phi}_{n} + \frac{1}{n_0/2} \frac{T^2}{n^2} {\bm \Phi}_{n}^{\mathrm{T}}{\bm K}_X {\bm \Phi}_{n} \right) \\
			\overset{(c)}{=} & \frac{1}{2} \lim_{n\rightarrow \infty}\log \det \left({\bm I}_n +\frac{1}{n_0/2} \mathrm{diag}(\lambda_1, \cdots, \lambda_n) \right) \\
			= & \frac{1}{2} \sum_{k=1}^{+\infty}{\log\left(1+\frac{\lambda_k}{n_0/2} \right)},
		\end{aligned}
		\label{eq_proof_source_expansion_awgn}
	\end{equation}
	where $(b)$ comes from sandwiching the determinant in the bracket with both the asymptotically orthogonal matrix $\sqrt{T/n}{\bm \Phi}_n$ on the left and its transpose on the right, and $(c)$ comes from plugging \eqref{eq_phi_n_orthogonal} and \eqref{eq_diagonalize_Kx} into the previous step.

\section*{Appendix B \\ Proof Of Lemma~\ref{lemma_1}}
	Tracing both left and right hand sides of \eqref{eq_diagonalize_Kx} and let $n\rightarrow \infty$, we obtain
	\begin{equation}
		\begin{aligned}
			\mathrm{tr}(M) & = \lim_{n\rightarrow \infty}\mathrm{tr}\left(\frac{T^2}{n^2}{\bm \Phi}_n^{\mathrm{T}} {\bm K}_X {\bm \Phi}_n \right) \\
			& = \lim_{n\rightarrow \infty}\left(\frac{T}{n} \times \mathrm{tr}\left(\frac{T}{n} {\bm \Phi}_n^{\mathrm{T}} {\bm K}_X {\bm \Phi}_n \right)\right) \\
			& = \lim_{n\rightarrow \infty} \left(\frac{T}{n} \mathrm{tr}\left({\bm K}_X \right)\right) \\
			& = \lim_{n\rightarrow \infty} \left(\frac{T}{n} \times nP\right) \\
			& = PT,\\
		\end{aligned}
	\end{equation}
	which completes the proof of {\bf Lemma \ref{lemma_1}}.

\section*{Appendix C \\ Proof Of Theorem \ref{th_2}}
	Plugging \eqref{eq_Csh_example2} into the right-hand side of \eqref{eq_exceed_shannon_special}, and differentiate both sides w.r.t $P$. Notice that if $P=0$, then both sides of \eqref{eq_exceed_shannon_special} are equal to 0. Thus, we only need to prove that the derivative of left-hand side is strictly larger than that of right-hand side within a small interval $P\in (0,\delta)$:
	\begin{equation}
		\frac{1}{2}\sum_{k=1}^{+\infty}{\left(\frac{1}{1+\frac{2\lambda_k}{n_0}}\frac{2\lambda_k}{n_0P}\right)} > \frac{T}{n_0}\frac{1}{\sqrt{1+4P/(n_0\alpha)}}.
		\label{eq_exceed_shannon_special_1}
	\end{equation}
	Multiply both sides of \eqref{eq_exceed_shannon_special_1} by $n_0$ and define $\mu_k:=\lambda_k/(PT)$, and then from {\bf Lemma \ref{lemma_1}} we obtain $\sum_k{\mu_k}=1$. In this way,  \eqref{eq_exceed_shannon_special_1} is equivalent to
	\begin{equation}
		\sum_{k=1}^{+\infty}{\frac{\mu_k}{1+\frac{2\lambda_k}{n_0}}} > \frac{1}{\sqrt{1+4P/(n_0\alpha)}}.
		\label{eq_exceed_shannon_special_2}
	\end{equation}
	Since $\varphi(x):=1/(1+2x/n_0)$ is convex on $(0,+\infty)$, by applying Jensen's inequality to the left-hand side  of \eqref{eq_exceed_shannon_special_2}, we only need to prove that
	\begin{equation}
		\frac{1}{1+\frac{2}{n_0}\sum_k{\lambda_k\mu_k}} > \frac{1}{\sqrt{1+4P/(n_0\alpha)}}.
		\label{eq_after_jensen}
	\end{equation}
	From the definition of $\mu_k$ we can derive that  $\lambda_k \mu_k = \lambda_k^2/(PT)$. So we go on to calculate $\sum_k{\lambda_k^2}$. That is equivalent to calculate $\mathrm{tr}(M^2)$, where $M^2$ corresponds to the integral kernel:
	\begin{equation}
		K_{M^2}(t_1, t_2) := \int_{0}^{T}{P^2\exp(-\alpha\lvert t_1-s\rvert)\exp(-\alpha\lvert t_2-s\rvert)\mathrm{d}s}.
	\end{equation}
	Evaluating the kernel $K_{M^2}$ on the diagonal $t=t_1=t_2$ yields
	\begin{equation}
		\begin{aligned}
			K_{M^2}(t,t) & =P^2\int_0^{T}{\exp(-2\alpha\lvert t-s \rvert)\mathrm{d}s}, \\
			& = \frac{P^2}{2\alpha}(2-\exp(-2\alpha t)-\exp(-2\alpha (T-t))). \\
		\end{aligned}
		\label{eq_exceed_shannon_special_2_1}
	\end{equation}
	Integrating this kernel on the diagonal of $[0,T]^2$ gives $\sum_{k}{\lambda_k^2}=\mathrm{tr}(M^2)$:
	\begin{equation}
		\begin{aligned}
			\sum_k{\lambda_k^2} & = \frac{P^2}{2\alpha}\int_{0}^{T}{\left(2-e^{-2\alpha t}-e^{-2\alpha (T-t)} \right) \mathrm{d}t}, \\
			& = \frac{P^2}{\alpha}\left(T - \frac{1}{2\alpha}(1-e^{-2\alpha T})\right). \\
		\end{aligned}
		\label{eq_exceed_shannon_special_3}
	\end{equation}
	By substituting \eqref{eq_exceed_shannon_special_3} into \eqref{eq_after_jensen}, we just need to prove that
	\begin{equation}
		\sqrt{1+4P/(n_0\alpha)} > 1+\frac{2P}{n_0\alpha}\left(1-\frac{1-e^{-2\alpha T}}{2\alpha T}\right).
		\label{eq_exceed_shannon_special_4}
	\end{equation}
	Define the dimensionless number $x=2P/(n_0\alpha)$. Since the function $\psi(x):=(1-\exp(-x))/x$ is strictly positive and less than 1 at $x>0$, we can conclude that, there exists a small positive $\delta>0$ such that \eqref{eq_exceed_shannon_special_4} holds for $0<P<\delta$. The number $\delta$ can be chosen as
	\begin{equation}
		\delta = \frac{n_0\alpha\psi(2\alpha T)}{(1-\psi(2\alpha T))^2} > 0,
	\end{equation}
	which implies that \eqref{eq_exceed_shannon_special_1} holds for any $0<P<\delta$. Thus, integrating \eqref{eq_exceed_shannon_special_1} on both sides from $p=0$ to $p=P, P<\delta$ gives rise to the conclusion \eqref{eq_exceed_shannon_special}, which completes the proof of {\bf Theorem \ref{th_2}}.

\section*{Appendix D \\ Proof Of Theorem \ref{th_3}}
	Differentiating the finite-time capacity expression for $I(T)$, i.e. \eqref{eq_th_1}, with respect to $T$, then we obtain
	\begin{equation}
		\frac{\partial I(T)}{\partial T} |_{T=0^+} = \lim_{T\rightarrow 0^+}\frac{1}{2}\sum_{k=1}^{+\infty}{\frac{1}{(n_0/2)(1+2\lambda_k/n_0)}\frac{\partial \lambda_k}{\partial T}},
		\label{eq_proof_th3_1}
	\end{equation}
	where $\lim_{T\rightarrow 0^+}\sum_{k=1}^{+\infty}\frac{\partial \lambda_k}{\partial T} $, by applying {\bf Lemma~\ref{lemma_1}}, can be expressed as
	\begin{equation}
		\begin{aligned}
			\lim_{T\rightarrow 0^+}\sum_{k=1}^{+\infty}\frac{\partial \lambda_k}{\partial T} & = \lim_{T\rightarrow 0^+}\frac{\partial}{\partial T} \sum_{k=1}^{+\infty}{\lambda_k} \\
			& = \frac{\partial}{\partial T} (PT) |_{T=0^+}\\
			& = P.
		\end{aligned}
	\label{eq_proof_th3_2}
	\end{equation}
	Since $\omega_k \uparrow +\infty$ as $T \downarrow 0^+$, we can safely conclude that $\lambda_k \downarrow 0^+$. From Dirichlet's test, the series in \eqref{eq_proof_th3_1} converges uniformly. Thus, by interchanging the infinite sum and the limit operation, we obtain
	\begin{equation}
		\begin{aligned}
			\frac{\partial I(T)}{\partial T} |_{T=0^+} & = \frac{1}{2} \sum_{k=1}^{+\infty}{\lim_{T\rightarrow 0^+}\left(\frac{1}{(n_0/2)(1+2\lambda_k/n_0)} \frac{\partial \lambda_k}{\partial T} \right) } \\
			& = \frac{1}{n_0} \sum_{k=1}^{+\infty}\frac{\partial \lambda_k}{\partial T} |_{T=0^+}\\
			& = \frac{1}{n_0} \lim_{T\rightarrow 0^+} \sum_{k=1}^{+\infty}\frac{\partial \lambda_k}{\partial T}\\
			& = \frac{P}{n_0},
		\end{aligned}
		\label{eq_proof_th3_3}
	\end{equation}
	which completes the proof of {\bf Theorem \ref{th_3}}.

\section*{Appendix E \\ The Evaluation of The Improper Integral \eqref{eq_Csh_example2}}
	Define the improper inegral with parameters $P>0$ and $\alpha>0$:
	\begin{equation}
		J(P, 
		\alpha):=\int_{-\infty}^{+\infty}{\log\left(1+\frac{P}{\alpha^2+\omega^2}\right){\rm
		 d}\omega}.
	\end{equation}
	By taking the partial derivative of $J(P,\alpha)$ with respect to $P$, we obtain
	\begin{equation}
		\frac{\partial J(P,\alpha)}{\partial P} = \int_{-\infty}^{+\infty}{ 
		\frac{1}{\omega^2 + (\alpha^2+P)} {\rm d}\omega}.
		\label{eq_proof_integral_1}
	\end{equation}
	Note that the analytic function defined as
	\begin{equation}
		f(z):=\frac{1}{z^2+(\alpha^2+P)},
	\end{equation}
	has residual
	\begin{equation}
		{\rm Res}\left[ f(z), z=z_p\right] = \frac{1}{2{\rm i}\sqrt{\alpha^2+P}},
	\end{equation}
	at pole $z_p={\rm i}\sqrt{\alpha^2+P}$ in the upper half-plane, thus the integral in \eqref{eq_proof_integral_1} can be evaluated by the residual theorem:
	\begin{equation}
		\begin{aligned}
			\frac{\partial J(P,\alpha)}{\partial P} & = 2\pi {\rm i}{\rm Res}\left[ f(z), z={\rm i}\sqrt{\alpha^2+P}\right]\\
			& = \frac{\pi}{\sqrt{\alpha^2+P}}.
		\end{aligned}
		\label{eq_proof_integral_2}
	\end{equation}
	Since $J(0,\alpha)\equiv 0$, by integrating \eqref{eq_proof_integral_1} with respect to $P$ from $0$ to $P$ yields
	\begin{equation}
		\begin{aligned}	
			J(P,\alpha) & = \int_{0}^{P} \frac{\pi}{\sqrt{\alpha^2+p}}{\rm d}p\\
			& = 2\pi \sqrt{\alpha^2+p} |_{p=0}^{P} \\
			& = 2\pi \left(\sqrt{\alpha^2+P}-\alpha \right).
		\end{aligned}
	\label{eq_proof_integral_3}
	\end{equation}
	Then the integral in \eqref{eq_Csh_example2} can be calculated by setting $P \leftarrow 4P\alpha/n_0$ and $\alpha \leftarrow \alpha$ in \eqref{eq_proof_integral_3}:
	\begin{equation}
		\begin{aligned}
			& \frac{1}{4\pi}\int_{-\infty}^{+\infty}{\log\left(1+\frac{\frac{2P\alpha}{\alpha^2+\omega^2}}{n_0/2}
			 \right)\mathrm{d}\omega} \\ 
			& = \frac{2\pi}{4\pi}\left( \sqrt{\alpha^2+\frac{4P\alpha}{n_0}} - \alpha \right) \\
			& = \frac{1}{2}\left( \sqrt{\alpha^2+\frac{4P\alpha}{n_0}} - \alpha \right),
		\end{aligned}
	\end{equation}
	which completes the proof.
	
\bibliography{IEEEabrv, refs}

\begin{thebibliography}{10}
\providecommand{\url}[1]{#1}
\csname url@samestyle\endcsname
\providecommand{\newblock}{\relax}
\providecommand{\bibinfo}[2]{#2}
\providecommand{\BIBentrySTDinterwordspacing}{\spaceskip=0pt\relax}
\providecommand{\BIBentryALTinterwordstretchfactor}{4}
\providecommand{\BIBentryALTinterwordspacing}{\spaceskip=\fontdimen2\font plus
\BIBentryALTinterwordstretchfactor\fontdimen3\font minus
  \fontdimen4\font\relax}
\providecommand{\BIBforeignlanguage}[2]{{%
\expandafter\ifx\csname l@#1\endcsname\relax
\typeout{** WARNING: IEEEtran.bst: No hyphenation pattern has been}%
\typeout{** loaded for the language `#1'. Using the pattern for}%
\typeout{** the default language instead.}%
\else
\language=\csname l@#1\endcsname
\fi
#2}}
\providecommand{\BIBdecl}{\relax}
\BIBdecl

\bibitem{Shannon}
C.~E. Shannon, ``A mathematical theory of communication,'' \emph{The Bell Syst.
  Techni. J.}, vol.~27, no.~3, pp. 379--423, Jul. 1948.

\bibitem{Sampling1967Landau}
H.~Landau, ``Sampling, data transmission, and the nyquist rate,'' \emph{Proc.
  {IEEE}}, vol.~55, no.~10, pp. 1701--1706, Oct. 1967.

\bibitem{Nyquist1928CertainTopics}
H.~Nyquist, ``Certain topics in telegraph transmission theory,'' \emph{Trans.
  American Institute of Electrical Engineers}, vol.~47, no.~2, pp. 617--644,
  Apr. 1928.

\bibitem{BW1976Slepian}
D.~Slepian, ``On bandwidth,'' \emph{Proc. {IEEE}}, vol.~64, no.~3, pp.
  292--300, Mar. 1976.

\bibitem{Cover1999ElementsInfTheory}
T.~M. Cover, \emph{Elements of information theory}.\hskip 1em plus 0.5em minus
  0.4em\relax John Wiley \& Sons, 1999.

\bibitem{donoho2006compressedsensing}
D.~L. Donoho, ``Compressed sensing,'' \emph{{IEEE} Trans. Inf. Theory},
  vol.~52, no.~4, pp. 1289--1306, Apr. 2006.

\bibitem{mercer1909functions}
J.~Mercer, ``Functions of positive and negative type and their connection with
  the theory of integral equations,'' \emph{Philos. Trans. Royal Soc.}, vol.
  209, pp. 4--415, 1909.

\bibitem{zhu2007operator}
K.~Zhu, \emph{Operator theory in function spaces}.\hskip 1em plus 0.5em minus
  0.4em\relax American Mathematical Soc., 2007, no. 138.

\bibitem{GPSampling1957BA}
A.~Balakrishnan, ``A note on the sampling principle for continuous signals,''
  \emph{IRE Trans. Inf. Theory}, vol.~3, no.~2, pp. 143--146, Jun. 1957.

\bibitem{MercerExpansion1955Barrett}
J.~Barrett and D.~Lampard, ``An expansion for some second-order probability
  distributions and its application to noise problems,'' \emph{IRE Trans. Inf.
  Theory}, vol.~1, no.~1, pp. 10--15, Mar. 1955.

\bibitem{Kelkar1983parsevaltheorem}
S.~S. Kelkar, L.~L. Grigsby, and J.~Langsner, ``An extension of parseval's
  theorem and its use in calculating transient energy in the frequency
  domain,'' \emph{{IEEE} Trans. Ind. Electron.}, vol. IE-30, no.~1, pp. 42--45,
  Feb. 1983.

\bibitem{brislawn1988kernels}
C.~Brislawn, ``Kernels of trace class operators,'' \emph{American Math. Soc.},
  vol. 104, no.~4, 1988.

\bibitem{Cai2020Eigenvalue}
D.~Cai and P.~S. Vassilevski, ``Eigenvalue problems for exponential-type
  kernels,'' \emph{Comput. Methods in Applied Math.}, vol.~20, no.~1, pp.
  61--78, Jan. 2020.

\bibitem{integralTable2014}
I.~S. Gradshteyn and I.~M. Ryzhik, \emph{Table of integrals, series, and
  products}.\hskip 1em plus 0.5em minus 0.4em\relax Academic press, 2014.

\end{thebibliography}

\end{document}